\documentclass[twocolumn,prl]{revtex4-2}

\usepackage{graphicx}
\usepackage{dcolumn}
\usepackage{bm}
\usepackage{multirow}
\usepackage{color}
\usepackage[normalem]{ulem}
\usepackage{amsmath}
\usepackage{gensymb}
\usepackage[colorlinks=true]{hyperref}
\usepackage{verbatim}

\usepackage{lineno}


\begin{document}


\title{Detecting Magnetic Phase Transitions in Ion-Irradiated CrSBr Through Resonant Raman Scattering}

\author{
Daria I. Markina$^{1}$$^{\ddagger}$,
Alison Pfister$^{1}$$^{\ddagger}$,
Priyanka Mondal$^{1}$,
Lukas Krelle$^{1}$,
Sai Shradha$^{1}$,
Regine von Klitzing$^{1}$,
Kseniia Mosina$^{2}$,
Zdenek Sofer$^{2}$,
Fangchao Long$^{3}$,
Ulrich Kentsch$^{3}$,
Shengqiang Zhou$^{3}$,
Bernhard Urbaszek$^{1}$*
}

\affiliation{\small $^1$Institute for Condensed Matter Physics, TU Darmstadt, Hochschulstraße 6-8, D-64289 Darmstadt, Germany}

\affiliation{\small $^2$Department of Inorganic Chemistry, University of Chemistry and Technology Prague, Technicka 5, 166 28 Prague 6, Czech Republic}

\affiliation{\small $^3$Helmholtz-Zentrum Dresden-Rossendorf, Institute of Ion Beam Physics and Materials Research, Dresden, Germany}
\affiliation{$^{\ddagger}$Authors contributed equally to this work}
\affiliation{\textcolor{magenta}{*bernhard.urbaszek@pkm.tu-darmstadt.de}}

\begin{abstract}
\textbf{Abstract --}
Controlling magnetic phases and accurately determining their transition temperatures are essential for the development of low-dimensional magnetic materials. Here, we demonstrate that He$^+$ ion irradiation provides a versatile route for engineering magnetic phases in layered CrSBr and establish temperature-dependent polarization-resolved Raman spectroscopy as a sensitive optical probe for identifying irradiation-induced magnetic phase transitions. We reveal that the magnetic response of the modified CrSBr is governed by both irradiation dose and crystal thickness. The temperature evolution of the Raman tensor elements resolves the antiferromagnetic transition in pristine CrSBr at $T_N \approx 132$ K as well as irradiation-induced magnetic transitions at $T_C \approx 105$–110 K and $T_D \approx 40$ K corresponding to ferromagnetic and defect-related magnetic phase transitions. Complementary magneto-optical measurements confirm the progressive suppression of antiferromagnetic order and the emergence of new defect-engineered magnetic phases, including pure ferromagnetic behavior at high irradiation doses. These findings establish irradiated CrSBr as a platform for controllable magnetic phase engineering while demonstrating polarization-resolved Raman spectroscopy as a rapid, non-destructive, and broadly applicable method for probing magnetic phase transitions in van der Waals magnets.
\end{abstract}

\maketitle

\section{Introduction} 
Determining and controlling magnetic phases is central to the development of two-dimensional magnetic materials for applications in spintronic and quantum technologies. Among 2D magnets, CrSBr has emerged as a particularly attractive platform for quantum applications owing to its strong magnetic, electronic, and optical anisotropy \cite{klein2024materials, yang2021triaxial, wang2023magnetic}. CrSBr is characterized by ferromagnetic ordering within each layer and antiferromagnetic coupling between adjacent layers \cite{telford2020layered, wang2023origin, long2023intrinsic}. It also exhibits a relatively high N\'eel temperature of approximately 132 K \cite{telford2020layered}.

While external magnetic fields can reversibly tune its magnetic state, permanent phase engineering requires modifying the interlayer exchange interaction through structural rearrangements or changes in interlayer spacing. Such modifications can be realized by electron or proton irradiation \cite{klein2022control, fassbender2012chemical}, chemical doping \cite{zhao2025doping, tabataba2024doping, sahu2026interplay}, charge transfer \cite{xie2023engineering}, cation intercalation \cite{shi2025controllable} or twisting layers \cite{mondal2026twist}. In particular, ion irradiation enables controlled introduction of defects without altering the material composition, providing a versatile route to engineer magnetic phases.

Reliable identification of the resulting magnetic phase transitions remains challenging, especially for two-dimensional materials. Established techniques, including SQUID magnetometry\cite{black2006measurements, buchner2018tutorial}, magnetic susceptibility measurements \cite{balanda2013ac, mugiraneza2022tutorial}, neutron and X-ray scattering \cite{shull1949detection, mcguire2015coupling, scheie2022spin}, and magneto-optical measurements \cite{krelle2025magnetic, wilson2021interlayer, smiertka2026distinct, kerr1877xliii, huang2017layer}, provide high precision, but often require specialized instrumentation or complex experimental procedures. Optical methods, like photoluminescence (PL) \cite{linhart2023optical, lin2024strong}, second harmonic generation (SHG) \cite{lee2021magnetic, ziebel2024crsbr}, or Raman scattering \cite{pawbake2023raman, torres2023probing, wdowik2025magneto} measurements, are particularly attractive because they are non-destructive, well suited for studying 2D magnets, and readily implemented. However, approaches based on phonon frequencies, linewidths, or intensities often provide only subtle signatures of magnetic ordering, limiting their sensitivity to complex or defect-induced magnetic phases.

Here, we demonstrate that temperature-dependent polarization-resolved Raman spectroscopy overcomes these limitations by probing changes in the Raman tensor associated with magnetic ordering. Using He$^+$-irradiated CrSBr as a model system, we show that this approach identifies both intrinsic and defect-engineered magnetic phase transitions with high sensitivity. Furthermore, we reveal that the engineered magnetic phases are governed not only by the irradiation dose but also by the sample thickness, establishing ion-irradiated CrSBr as a promising platform for tuning magnetism in layered materials.

\section{Results and discussion}
\subsection{Magnetic Phase Control in CrSBr}

In this work, we employ He$^+$ ion irradiation to control the magnetic phase of CrSBr from few-layer to bulk samples (Fig.~\ref{fig:fig1}a) without applying an external magnetic field. Previous studies \cite{long2023ferromagnetic, long2024rise} have shown that He$^+$ irradiation induces defect formation in CrSBr. The dominant defects are Frenkel pairs, which consist of a displaced lattice atom and the vacancy left behind, while the displaced atom occupies an interstitial position (Fig.~\ref{fig:fig1}b). Bromine vacancies were reported to exhibit the lowest defect formation energy \cite{torres2023probing, long2023ferromagnetic} and were directly identified using scanning tunneling microscopy \cite{klein2022sensing}. However, Sputtering Range of Ion in Matter (SRIM) simulations \cite{torres2023probing} indicate that ion irradiation is also expected to significantly increase the concentrations of sulfur and chromium vacancies. In addition, Long et al. \cite{long2023ferromagnetic} reported lower Frenkel pair formation energies for sulfur and bromine. Together, these observations suggest that all three atomic species constituting the CrSBr crystal lattice can form Frenkel pairs under He$^+$ irradiation.

The samples investigated in this study were irradiated with two different doses corresponding to the following ion fluences: (1) 3.8$\cdot$10$^{13}$–1.5$\cdot$10$^{14}$ cm$^{-2}$, referred to throughout the text as the single-dose condition, and (2) 7.6$\cdot$10$^{13}$–3.0$\cdot$10$^{14}$ cm$^{-2}$, referred to as the double-dose condition. The ion energy and fluence were selected to produce a nearly uniform displacement per atom (DPA) level of 0.01 and 0.02 throughout the entire thickness of the layer, consistent with the irradiation conditions reported by Long et al.\cite{long2023ferromagnetic}. 

To investigate the magnetic phases formed after irradiation, we employed temperature-dependent polarization-resolved Raman spectroscopy. In our previous work \cite{markina2026interplay}, we demonstrated that this technique provides insights into spin--phonon interactions and enables tracking of magnetic phase transitions. Figure~\ref{fig:fig1}c shows Raman scattering spectra of a double-dose sample measured at 4 K under 1.96 eV laser excitation with polarization along the \textit{a-} and \textit{b-} crystallographic axes. An excitation energy of 1.96 eV was chosen to provide near-resonant excitation conditions \cite{markina2026interplay}. The spectra retain the overall shape observed in pristine (non-irradiated) samples (Fig.\textcolor{red}{S1}). Similar to the pristine case, the three dominant optically active modes correspond to out-of-plane vibrations of A$_g$ symmetry and preserve pronounced anisotropy, exhibiting maximum intensity along the \textit{b-}axis (Fig.~\ref{fig:fig1}d).

It is worth noting that the second-order Raman scattering features observed between 400 and 800 cm$^{-1}$, associated with the establishment of intralayer ferromagnetic order \cite{pawbake2023raman}, become significantly suppressed after irradiation and therefore cannot be reliably used to analyze phase transitions. Meanwhile, the broad feature near the A$_g^3$ mode may originate from irradiation-induced defects \cite{torres2023probing, sahu2026interplay}. 
However, a similar peak is also present in the as-exfoliated pristine sample in the antiferromagnetic phase, disappearing above T$_N$ (Fig.\textcolor{red}{S1}). This peak could alternatively correspond to the B$_{3g}^3$ mode \cite{Chen2026CrSBrRaman}. 

\subsection{Raman Scattering Temperature Evolution}

Raman spectroscopy was performed with an angular resolution of 4 degrees over a wide temperature range from 4 to 200 K to investigate magnetic phase modifications in pristine and He$^+$-irradiated CrSBr flakes. To assess possible thickness-dependent effects, two regions with different thicknesses were measured: few-layer (8-9 nm thick) and bulk (30--90 nm thick) flakes. All measurements were carried out in the co-polarized geometry (Fig.\textcolor{red}{S2}) using 1.96 eV laser excitation to provide near-resonant excitation \cite{markina2026interplay, mondal2025raman}. The angular dependencies were acquired at each temperature, yielding complete polar plots (Fig.~\ref{fig:fig1}d).

As demonstrated in \cite{markina2026interplay}, these polar dependencies contain rich information that can be extracted by fitting the angular profiles using the following expression \cite{pimenta2021polarized}:
\begin{equation}
    \label{fit}
    I_{||}^{A_g}({\theta})=(a{\cdot}\text{cos}^2{\theta}+b{\cdot}\text{cos}{\phi}_{ab}{\cdot}\text{sin}^2{\theta})^2+b^2{\cdot}\text{sin}^4{\theta}{\cdot}\text{sin}^2{\phi}_{ab},
\end{equation}
where parameters $a$ and $b$ represent the Raman tensor amplitudes along the $a$- and $b$-axes, respectively, while the phase $\phi_{ab}$ accounts for the imaginary part of the Raman tensor elements arising from the interplay between virtual and real intermediate states.

Figure ~\ref{fig:fig2}a shows the temperature evolution of the $b$ component of the Raman tensor for a pristine sample (parameter $b$ obtained from the fitting by Eq.~\ref{fit}). The A$_g^3$ mode of the pristine sample exhibits a pronounced discontinuity near the N\'eel temperature (132 K), corresponding to the transition from the paramagnetic (PM) to the antiferromagnetic (AFM) phase (Fig.~\ref{fig:fig3}b). As explained in \cite{markina2026interplay}, this discontinuity originates from electron-assisted spin--phonon coupling and is observed consistently in CrSBr flakes with thicknesses ranging from 5 to 90 nm.

The irradiated samples exhibit similar discontinuities, but at different temperatures. In particular, two distinct dips appear at approximately 40 K (T$_D$) and 105--110 K (T$_C$) (Fig.~\ref{fig:fig2}b-e). We attribute these anomalies to magnetic phase transitions. The transition at 105--110 K most likely corresponds to the onset of magnetic ordering associated with spin alignment from the paramagnetic state to a modified antiferromagnetic (AFM) or ferromagnetic (FM) phase (Fig.~\ref{fig:fig3}c), in agreement with magnetic moment measurements reported in \cite{long2023ferromagnetic, long2024rise}. The second anomaly at 40 K indicates the emergence of an additional magnetic phase. Its origin could be related to ferromagnetic ordering of irradiation-induced magnetic defects, spin freezing caused by the suppression of spin fluctuations, or a crossover from XY to triaxial magnetic anisotropy, as discussed in previous studies \cite{lopez2022dynamic, telford2022coupling, klein2022sensing, ziebel2024crsbr, torres2023probing, pawbake2023raman, lin2024strong, wdowik2025magneto, telford2020layered}. Considering the dilute defect concentration and the fact that exchange interactions are effective over atomic length scales, which are shorter than the typical distance between defects, the precise nature of this transition remains an open question and requires further detailed investigation. Interestingly, signatures of both T$_N$ and T$_C$ were never observed simultaneously in the same sample.

The magnitude of the observed dips increases with irradiation dose but decreases with sample thickness (Fig.\ref{fig:fig2}). The former trend is expected, as stronger irradiation modifies the magnetic exchange interactions more significantly, resulting in a more pronounced Raman response. The thickness dependence can also be understood in terms of defect formation. This is evident from the calculated displacement per atom (DPA) profile (Fig.~\ref{fig:fig3}a), which estimates the depth distribution of energy transferred from the incident ions to the crystal lattice and, consequently, the spatial distribution of irradiation-induced defects. 
The DPA is significantly lower in the first few layers than deeper inside the crystal. This is because the displacement of crystal lattice atoms, caused by nuclear collisions between He$^+$ ions and target atoms, is generally greater when the incident He$^+$ ions have lower energy. Near the surface, the He$^+$ ions still possess relatively high energy, resulting in fewer nuclear collisions and, consequently, a lower DPA.
We propose that, although bulk flakes experience a stronger modification of the magnetic exchange interaction, their vibrational response could be partially suppressed due to the high density of irradiation-induced defects, resulting in less pronounced Raman anomalies in thicker flakes.

It is also noteworthy that, unlike the pristine sample, all three Raman modes exhibit a signature at T$_D$, while the A$_g^1$ and A$_g^3$ modes also display a discontinuity at T$_C$. Furthermore, for the sample exhibiting the strongest anomalies (the double-dose few-layer flake, Fig.~\ref{fig:fig2}e), the A$_g^2$ mode also develops a clear discontinuity at T$_C$. We attribute this mode-dependent behavior to irradiation-induced modifications of the electronic (or phononic) band structure. As different Raman modes in CrSBr couple selectively to different electronic states \cite{mondal2025raman}, they are expected to respond differently to these changes.

The phonon energies decrease monotonically with increasing temperature and show no discernible anomalies at the phase transition temperature in either the pristine or the irradiated samples (Fig. S3). In contrast to previous reports on CrSBr \cite{torres2023probing, wdowik2025magneto}, the temperature evolution of the Raman shifts does not provide signatures of the magnetic phase transitions. It thus precludes conclusions regarding the nature of the spin--phonon coupling. It is noteworthy that the phonon energies of all three A$_g$ modes decrease after irradiation, as shown in \cite{long2023ferromagnetic, long2024rise}. Furthermore, the decrease in phonon energy becomes more pronounced as the radiation dose increases. In contrast, the phonon energies are nearly independent of the sample thickness.

\subsection{Magnetic Phases Probing with Magneto-Optical Spectroscopy}

To investigate the formation of the modified magnetic phase induced by He$^+$ irradiation, we performed magneto-optical spectroscopy at 4.7 K using a three-axis vector magnet \cite{krelle2025magnetic}. An external magnetic field was applied along the $b$-axis. The pristine sample with a thickness of 5 nm exhibits the characteristic magnetic-field dependence of the PL energy, with a step-like redshift of approximately 13 meV occurring above the saturation field, B$^S_b$ = \text{--}0.37$\pm$0.01 T and 0.41$\pm$0.01 T, depending on the sweep direction (yellow dashed line, Fig.~\ref{fig:fig4}c). The measured values of the PL energy shift and saturation fields are in good agreement with previous reports \cite{krelle2025magnetic, wilson2021interlayer}.

In contrast, the PL spectra of the irradiated samples lose the fine structure characteristic of few-layer and bulk CrSBr (Fig.~\ref{fig:fig4}b) and instead exhibit a broad emission peak. The peak energy depends on both the irradiation dose and the sample thickness (Fig.~\ref{fig:fig4}a, b). 

We also investigated the excitation power dependence of the PL intensity in pristine and irradiated samples of different thicknesses using 1.96 eV excitation (Fig. S4). The extracted power-law exponent, $\alpha$, ranges from 0.916 to 1.021 for all samples, confirming a linear dependence of the PL intensity on the excitation power. Furthermore, no PL saturation is observed up to an excitation power of 1000~$\mu$W.

We find that the single-dose irradiated samples still exhibit magnetic-field-induced PL energy shifts for both thicknesses. The saturation fields are substantially reduced. In the few-layer region, the saturation fields decrease to B$^S_b$ = \text{--}0.19$\pm$0.02 T and 0.20$\pm$0.02 T, accompanied by a total PL shift of approximately 4 meV. In the bulk region, the saturation fields are further reduced to B$^S_b$ = \text{--}0.11$\pm$0.02 T and 0.09$\pm$0.02 T, with a PL shift of 3--4 meV (Fig.~\ref{fig:fig4}e,f). Both the saturation fields and the PL shifts are more than a factor of two smaller than those of the pristine sample. These results indicate that single irradiation does not lead to a complete transition into the FM phase. Instead, it substantially modifies the AFM state by altering the underlying magnetic exchange pathways.

In the double-dose irradiated sample, the few-layer region exhibits only barely discernible changes in the PL field sweep (Fig.~\ref{fig:fig4}g). Nevertheless, its zero-field PL energy is blueshifted by approximately 20 meV relative to the bulk region. In contrast, the bulk region exhibits no signature of a magnetic phase transition in the PL field dependence, consistent with a fully ferromagnetic state (Fig.~\ref{fig:fig4}h). The dependence of the total PL energy shifts and saturation magnetic fields on irradiation dose and thickness, including additional data points, is presented in Fig.~\ref{fig:fig4}d. The data clearly show that He$^+$ irradiation, even at relatively low doses, significantly affects the magnetic phase.

The observed thickness-dependent behavior can be understood in terms of the DPA profile (Fig.~\ref{fig:fig3}a). Since the uppermost layers experience lower defect density than the bulk, they undergo a weaker modification of the magnetic exchange interactions. As a result, the few-layer region retains a residual magnetic-field response, whereas the more heavily irradiated bulk region exhibits behavior consistent with complete ferromagnetic ordering.

\subsection{Mechanisms of Magnetic Transitions Detection}

To understand the sensitivity of polarization-resolved Raman scattering to irradiation-induced magnetic phase transitions, it is first necessary to consider how He$^+$ irradiation modifies the crystal structure and, consequently, the magnetic exchange interactions in CrSBr.

Monolayer CrSBr is intrinsically ferromagnetic. Its magnetic ground state is governed by the first-nearest-neighbor Cr--Br--Cr and Cr--S--Cr superexchange interactions, together with the second-nearest-neighbor Cr--S--Cr exchange pathways \cite{Xu2022StrongSpinPhonon}. Both sulfur and bromine atoms actively participate in mediating the intralayer superexchange interaction \cite{linhart2023optical}. In contrast, multilayer CrSBr exhibits antiferromagnetic interlayer coupling originating from the super-superexchange interaction between adjacent layers mediated predominantly by the $p$ orbitals of bromine atoms \cite{linhart2023optical}. Although the interlayer exchange interaction is approximately one order of magnitude weaker than the intralayer exchange \cite{klein2022control, bo2023calculated}, it determines the overall magnetic ordering of the crystal.

He$^+$ irradiation introduces Frenkel defects, where displaced atoms occupy interstitial positions within the van der Waals gap. These interstitial atoms form additional covalent bonds between adjacent layers \cite{klein2022control}, thereby modifying the interlayer exchange pathways. In particular, for chromium interstitials, the first-nearest-neighbor interlayer exchange interaction becomes mediated by orbital hybridization between the Cr interstitial and the inner Br atoms \cite{long2023ferromagnetic}. As a consequence, the antiferromagnetic interlayer coupling is progressively weakened and may eventually be overcome by ferromagnetic exchange, resulting in the irradiation-induced magnetic phases observed in this work.

The optical response of CrSBr is closely linked to its magnetic ordering. As demonstrated in our previous work \cite{markina2026interplay}, under near-resonant excitation ($E_L = 1.96$ eV), the transition into the ferromagnetic phase in the monolayer limit breaks time-reversal symmetry and lifts the degeneracy of excitonic states, giving rise to an exchange-field-induced Zeeman-like splitting of the excitonic resonances. Since Raman scattering proceeds through real or virtual electronic excitations, modifications of the electronic structure associated with magnetic ordering directly impact the Raman tensor components. Raman tensor components are expressed as \cite{yu2010fundamentals}:
%
\begin{multline}
     \label{Raman}
    R_{is}^{\mu}(E_L)\\
    =\sum_{n,n'}\frac{\langle 0|\mathbf e_s^*\cdot \hat{\mathbf{p}}|n'\rangle\langle n'|H_{el-ph}^{\mu}|n\rangle \langle n|\mathbf e_i\cdot \hat{\mathbf{p}}|0\rangle}{[E_L-E_{n}+i\Gamma_n][E_L-E_{ph}^{\mu}-E_{n'}+i\Gamma_{n'}]},
\end{multline}
where $E_{L}$ is the laser excitation energy, $E_{n}$ and $E_{n'}$ are the energies of the intermediate electronic states $|n\rangle$ and $|n'\rangle$, $\Gamma_n$ and $\Gamma_{n'}$ are their damping constants, $\hat{\bm p}$ is the momentum operator, and $H_{el-ph}^{\mu}$ is the electron--phonon interaction Hamiltonian for the phonon mode $\mu$ with energy $E_{ph}^{\mu}$. The indices $i$ and $s$ denote the polarization states of the incident and scattered photons, respectively.

Since magnetic ordering modifies the electronic band structure (i.e., all terms in Eq.~\ref{Raman} involving the intermediate states $|n\rangle$ and $|n'\rangle$), excitonic resonances, and the electron--phonon coupling ($\langle n'|H_{el-ph}^{\mu}|n\rangle$), all three quantities become temperature dependent across a magnetic phase transition. As a result, the Raman tensor components exhibit characteristic discontinuities at the transition temperatures \cite{markina2026interplay}, consistent with the experimental observations in Fig.~\ref{fig:fig2}. The pronounced anomalies observed for the $A_g^1$ and $A_g^3$ modes indicate a particularly strong sensitivity to the magnetic rearrangement, whereas the weaker response of the $A_g^2$ mode suggests that it couples to different intermediate electronic states, $|n\rangle$ and $|n'\rangle$.

In conclusion, we demonstrate two complementary advances. First, He$^+$-irradiated CrSBr emerges as a promising platform for defect-engineered magnetism, exhibiting irradiation-induced magnetic phase transitions near 40 K and 105–110 K. The higher-temperature transition is associated with a change in spin alignment from the paramagnetic state to a modified antiferromagnetic (AFM) or ferromagnetic (FM) phase. The precise nature of the lower-temperature transition, which is attributed to irradiation-induced defects, remains an open question and requires further investigation. Our results show that the resulting magnetic behavior is governed not only by the irradiation dose but also by the sample thickness, providing an additional degree of control over the engineered magnetic phases.
Second, we establish temperature-dependent, polarization-resolved Raman spectroscopy as a powerful, non-destructive optical approach for detecting these magnetic phase transitions. We demonstrate that this technique sensitively tracks changes in magnetic order resulting from irradiation-induced modifications of the magnetic exchange interactions.
Together, these findings open new possibilities for engineering magnetic states and materials with tunable functionalities.

\textbf{METHODS}\\
\indent\textbf{Sample Fabrication}\\
Bulk CrSBr crystals were fabricated through chemical vapor transport \cite{klein2022control}. The samples were prepared through mechanical exfoliation onto SiO$_2$/Si substrates with an 85 nm SiO$_2$ layer.\\
\indent\textbf{Atomic Force Microscopy}\\
To probe the topology and thickness of the CrSBr flakes atomic force microscopy measurements were performed at room temperature on a Cypher AFM (Asylum Research/Oxford Instruments, Wiesbaden, Germany). Height images of CrSBr flakes were obtained in AC tapping mode using the cantilever AC160TSA-R3 (300 kHz, 26 N/m, 7 nm tip radius). Images were post-processed with the in-built software features of IGOR 6.38801 (16.05.191, Asylum Research, Santa Barbara, CA, USA).\\
\indent\textbf{He$^+$ Irradiation}\\
For ion irradiation, He beam with energies of 8 keV and 1 keV were used to produce a relatively homogeneous defect profile throughout the entire thickness of the flakes. In the first irradiation run, the doses were $1.5\times10^{14}$ cm$^{-2}$ (8 keV) and $3.8\times10^{13}$ cm$^{-2}$ (1 keV), respectively. In the second run, the samples were irradiated with twice these doses. Throughout the manuscript, these samples are referred to as the single-dose and double-dose samples, respectively. Under these irradiation conditions, the double-dose samples are expected to have a defect density comparable to that of the ferromagnetic sample reported in \cite{long2023ferromagnetic}.\\
\indent\textbf{Displacement Per Atom (DPA) Calculation}\\
To estimate the DPA for different ion fluences, we performed Monte Carlo simulations of the collisions between He ions and the target atoms. The Stopping and Range of Ions in Matter simulations were carried out using the "Quick Calculation of Damage" option. A default displacement energy of 25 eV was assumed for all elements, as, to the best of our knowledge, no experimental or theoretical values have been reported for CrSBr. These simulations are intended only to provide a qualitative estimate of the upper limit of the damage and its depth profile.\\
\indent\textbf{Magneto-Optical Spectroscopy}\\
Optical spectroscopy was carried out in a home-built confocal setup for magneto-optical spectroscopy \cite{shree2021guide}. The sample was placed inside a closed-cycle cryostat (attocube systems, AttoDry 1000XL) equipped with a vector magnet (z axis, solenoid, maximum field = 5 T; x/y axis, split coil, maximum field = 2 T). We used low-temperature piezopositioners (attocube systems, ANPx101 and ANPz102) to position the sample with respect to a low-temperature apochromatic objective. PL measurements were performed in backscattering geometry at a sample temperature of 4.7 K. The signal was dispersed inside a Czerny–Turner spectrograph (Teledyne Princeton Instruments, SpectraPro HRS-500) and detected by a CCD camera (Teledyne Princeton Instruments, Pylon BRexcelon 100). For excitation, we used a HeNe laser (Thorlabs, HNL210LB). PL emission was detected along the $b$-axis. PL measurements were performed at an excitation power of 100 $\mu$W. Magnetic-field-dependent measurements were performed by initializing the CrSBr sample in the FM state, ramping the magnet to -0.6 T (-0.3 T) for the pristine (irradiated) sample, followed by a sweep to 0.6 T (0.3 T) and subsequent inversion of the sweep direction to obtain a full hysteresis.\\
\indent\textbf{Raman Spectroscopy}\\
Raman polarization-dependent measurements were performed using the same type of home-built spectroscopy setup as for magneto-optical spectroscopy. The sample was placed inside a closed-cycle cryostat (Attocube systems, AttoDry 800). For above-band-gap excitation HeNe laser (E$_L$ = 1.96 eV, Thorlabs) was used. The incident light was polarized using a Glan-Laser prism and then rotated with a Liquid Crystal Rotator placed in front of the objective (50x, NA = 0.7, CryoGlass Optics) on a low-temperature piezo-positioners (Attocube systems, ANPx101 and ANPz102) to position the sample with respect to the objective. The objective in the backscattering geometry focused the laser beam on the sample surface at normal incidence. The spot size diameter was of the order of the wavelength used, confirmed experimentally for different wavelengths by scanning the reflection signal over a metal stripe. Spectral purity of the excitation light was ensured by using MaxLine filter (Semrock). The Raman signal was collected using the same objective and sent via free space to the spectrometer (Teledyne Princeton Instruments IsoPlane300) coupled to a thermoelectrically cooled CCD (Teledyne Blaze 400HRX). The collected signal was dispersed with a 1200 lines/mm diffraction grating. A spectral blocking of the Rayleigh was made with a long-pass filter (Verona Long-pass Raman Edge Filter, Semrock). The average power for all measurements was maintained in the range of 200-400 $\mu$W. All measurements were performed in a co-polarization configuration provided by placing a Glan-Laser polarizer before the spectrometer with the optical axis parallel to the Glan-Laser polarizer in the excitation path. \\


\textbf{Data availability}\\
The datasets generated and analyzed during the current study are publicly available at the following link: \href{https://tudatalib.ulb.tu-darmstadt.de/handle/tudatalib/5541}{https://tudatalib.ulb.tu-darmstadt.de/handle/tudatalib/5541}.\\

 \textbf{Acknowledgements}\\
Z.S. was supported by ERC-CZ program (project LL2101) from Ministry of Education Youth and Sports (MEYS) and by the Advanced Multiscale Materials for Key Enabling Technologies project, supported by the Ministry of Education, Youth, and Sports of the Czech Republic. Project No. CZ.02.01.01/00/22$\_$008/0004558, Co-funded by the European Union. Helium ion irradiation was performed at the Ion Beam Center of Helmholtz-Zentrum Dresden-Rossendorf. We thank Anton L\"ogl, Gang Wang, and Mikhail M. Glazov for technical assistance and scientific discussions.\\

\textbf{Author contributions} 
K.M. and Z.S. grew the bulk CrSBr crystals. D.I.M. and P.M. fabricated few-layer CrSBr samples for optical spectroscopy. F.L., U.K., and S.Z. performed He$^+$ ion irradiation of CrSBr samples and provided DPA calculations. D.I.M, A.P., and L.K. performed optical spectroscopy measurements. S.S installed the cryostat system. P.M., A.P., and R.v.K. performed and interpreted AFM measurements. D.I.M., A.P., S.Z. and B.U. analyzed the optical spectra.
All authors discussed the results. B.U., S.Z. and D.I.M. suggested the experiments and supervised the project. D.I.M., A.P., and B.U. wrote the manuscript with input from all the authors. All authors have read and approved the manuscript.\\

\textbf{Competing interests}: The authors declare no competing interests.\\


\begin{figure*}[t]
\includegraphics[width=1\linewidth]{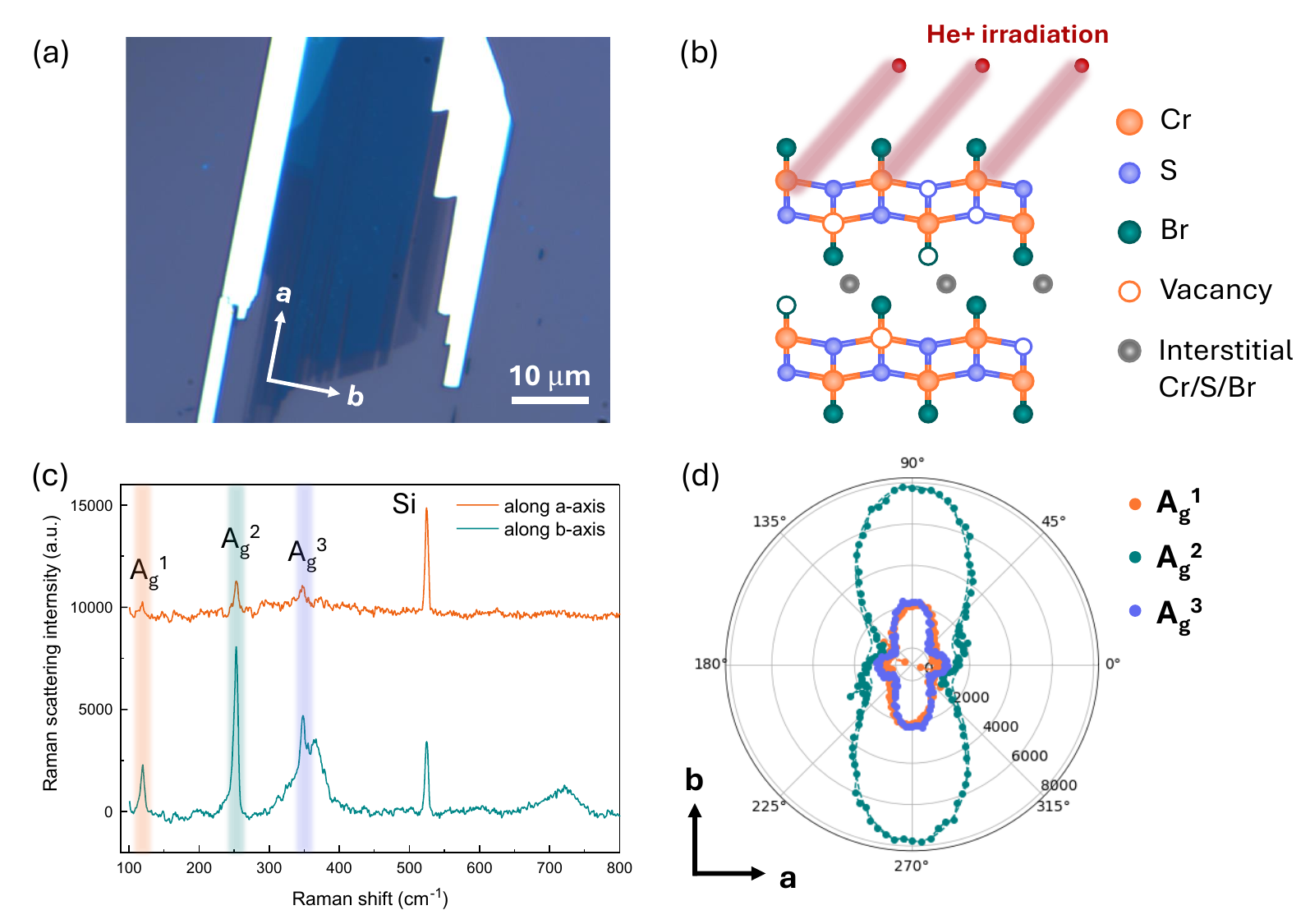}
\caption{\label{fig:fig1} \textbf{Control of the magnetic phase in CrSBr.} (a) White light image of exfoliated flake after irradiation with a single dose. (b) Schematic of the He$^+$ irradiation of CrSBr process. (c) Raman scattering spectra of a double-dose irradiated sample in the few-layer region at 4 K polarized along the $a$-axis (orange line) and $b$-axis (green line). (d) Polar plot of the Raman scattering intensity as a function of the polarization rotation angle for the A$_g^1$ mode (orange), A$_g^2$ mode (teal), and A$_g^3$ mode (blue) for a double-dose irradiated sample under excitations of 1.96 eV (left column). The polarization angle is measured relative to the $a$-axis (0$^\circ$). The radial axis indicates signal intensity with the center corresponding to 0 counts.}
\end{figure*}

\begin{figure*}[t]
\includegraphics[width=1\linewidth]{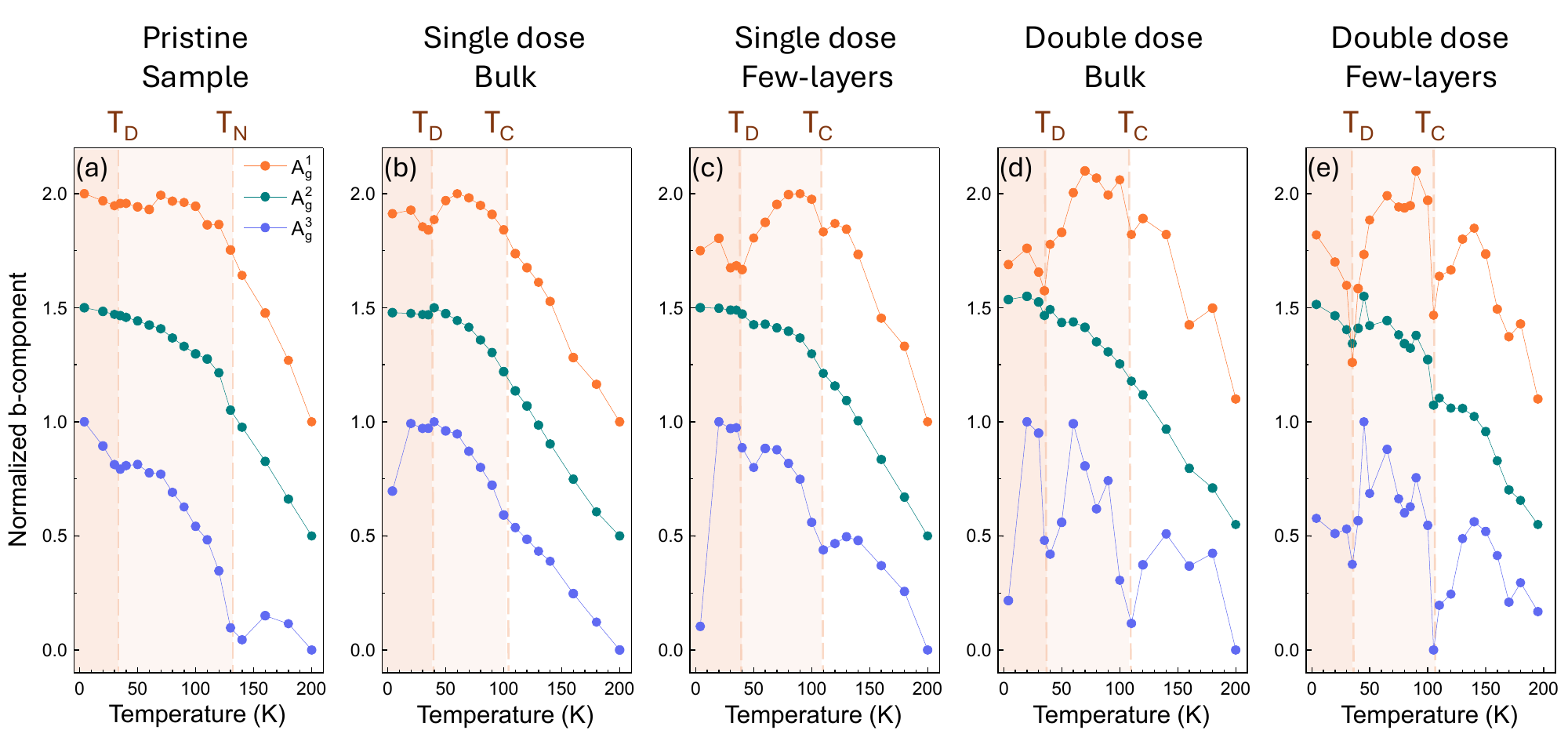}
\caption{\label{fig:fig2} \textbf{Temperature-dependent $b$-component of Raman tensor of the three major Raman modes under 1.96 eV excitation for different sample irradiation doses and thicknesses.} Plots display the normalized $b$-component of the Raman tensor of A$_g^1$ (orange dotted line), A$_g^2$ (teal dotted line), and A$_g^3$ modes (blue dotted line) for (a) pristine few-layer, (b, c) single-dose irradiated bulk and few-layer, and (d,e) double-dose irradiated bulk and few-layer samples. The orange-shaded region marks the different magnetic phases, the orange dashed line indicates the N\'eel (T$_N$ = 132 K), Curie (T$_C$ = 105-110 K), or defect-related (T$_D$ = 40 K) phase transition temperatures. The $b$ component was extracted from fits of the polarization dependence using Eq.~\eqref{fit}.}
\end{figure*}

\begin{figure*}[t]
\includegraphics[width=1\linewidth]{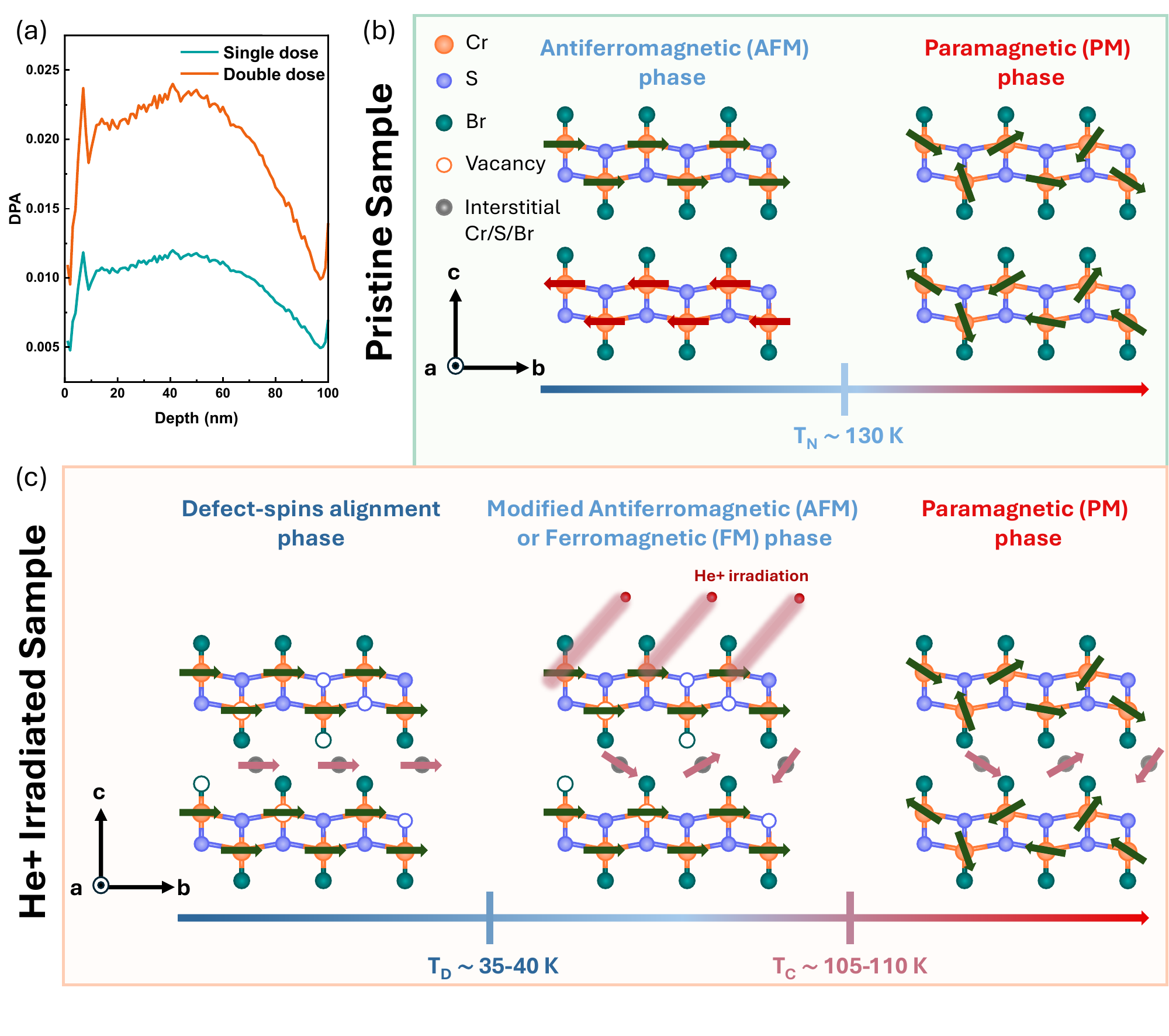}
\caption{\label{fig:fig3} \textbf{Magnetic phase transitions caused by the He$^+$ irradiation of CrSBr.} (a) Depth-dependent DPA for single- (green curve) and double-dose (orange curve) irradiations. Schematic of the phase transitions in (b) pristine sample and (c) He$^+$ irradiated sample.}
\end{figure*}

\begin{figure*}[t]
\includegraphics[width=0.85\linewidth]{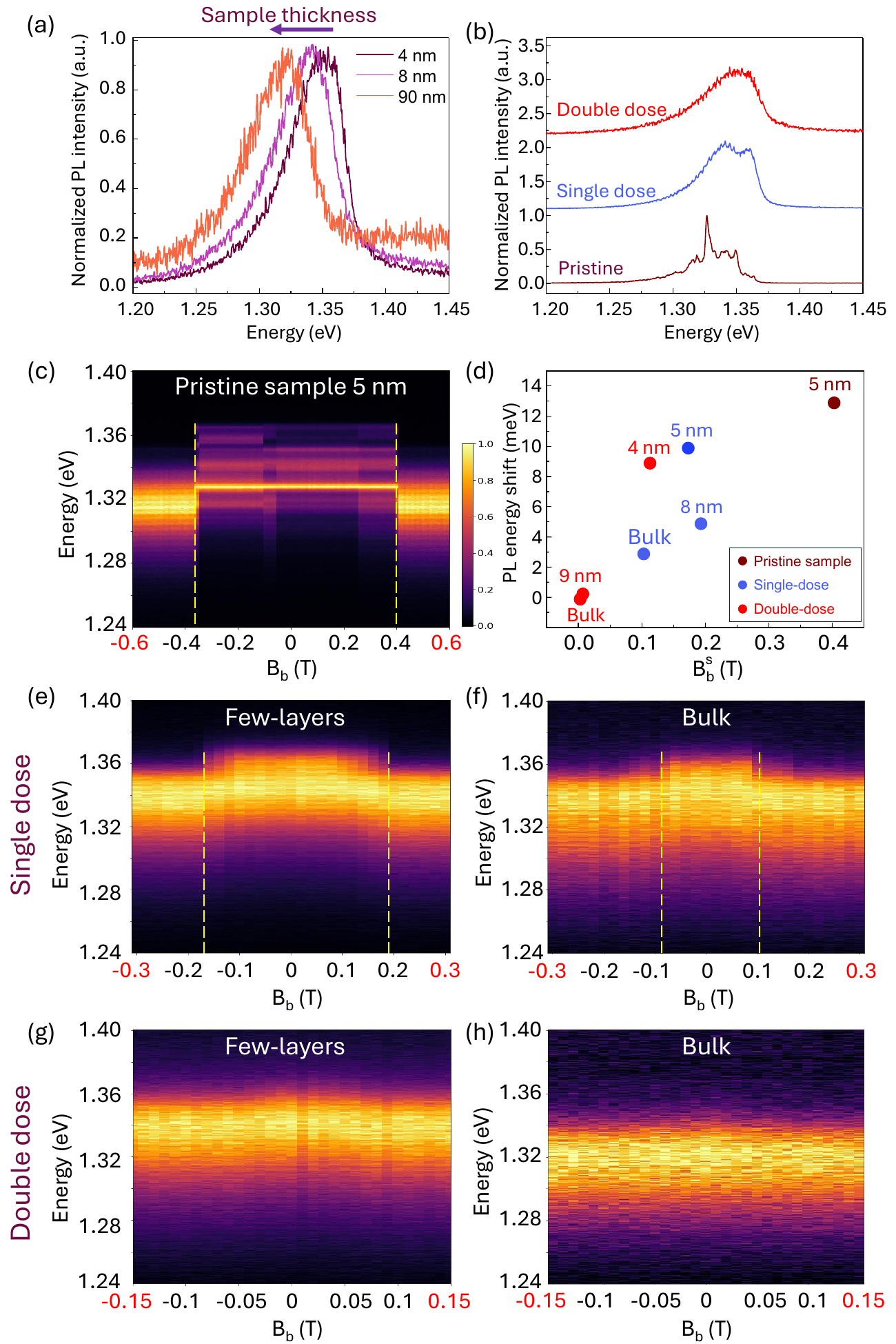}
\caption{\label{fig:fig4} \textbf{Probing of the magnetic phases with magneto-optical spectroscopy}. PL spectra of the (a) double-dose irradiated sample of different thicknesses at $B_b=0$ and (b) samples of 5 nm thickness irradiated with different doses at $B=0$. (c) PL magnetic field sweeps of a pristine few-layer sample along the $b$- crystallographic axis. (d) PL energy shift at maximum intensity versus external magnetic field for the samples of different thicknesses and irradiation doses. PL magnetic field sweeps of (e, f) single-dose and (g, h) double-dose irradiated samples of few-layer (left column) and the bulk (right column) thicknesses. Yellow dashed lines are a guide for the eye for the saturation fields. }
\end{figure*}

\renewcommand{\thefigure}{S\arabic{figure}}
\setcounter{figure}{0}

\begin{figure*}
\includegraphics[width=1\linewidth]{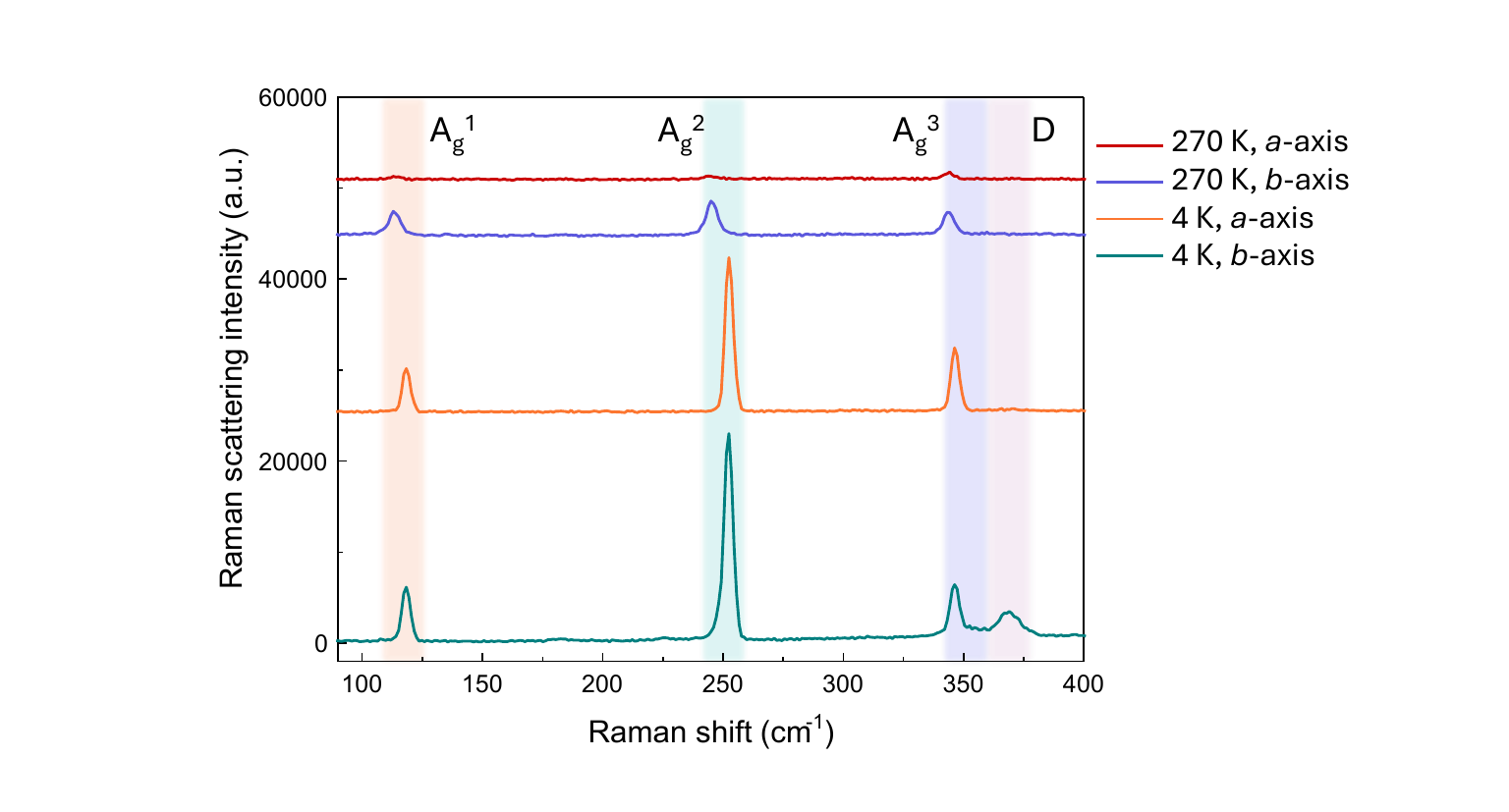}
\caption{\textbf{Raman scattering of the pristine sample.} Typical Raman spectra for pristine CrSBr of 5 nm thick. Excitation energy is 1.96 eV. }
\label{fig:fig1} 
\end{figure*}
\newpage

\begin{figure*}
\includegraphics[width=0.8\linewidth]{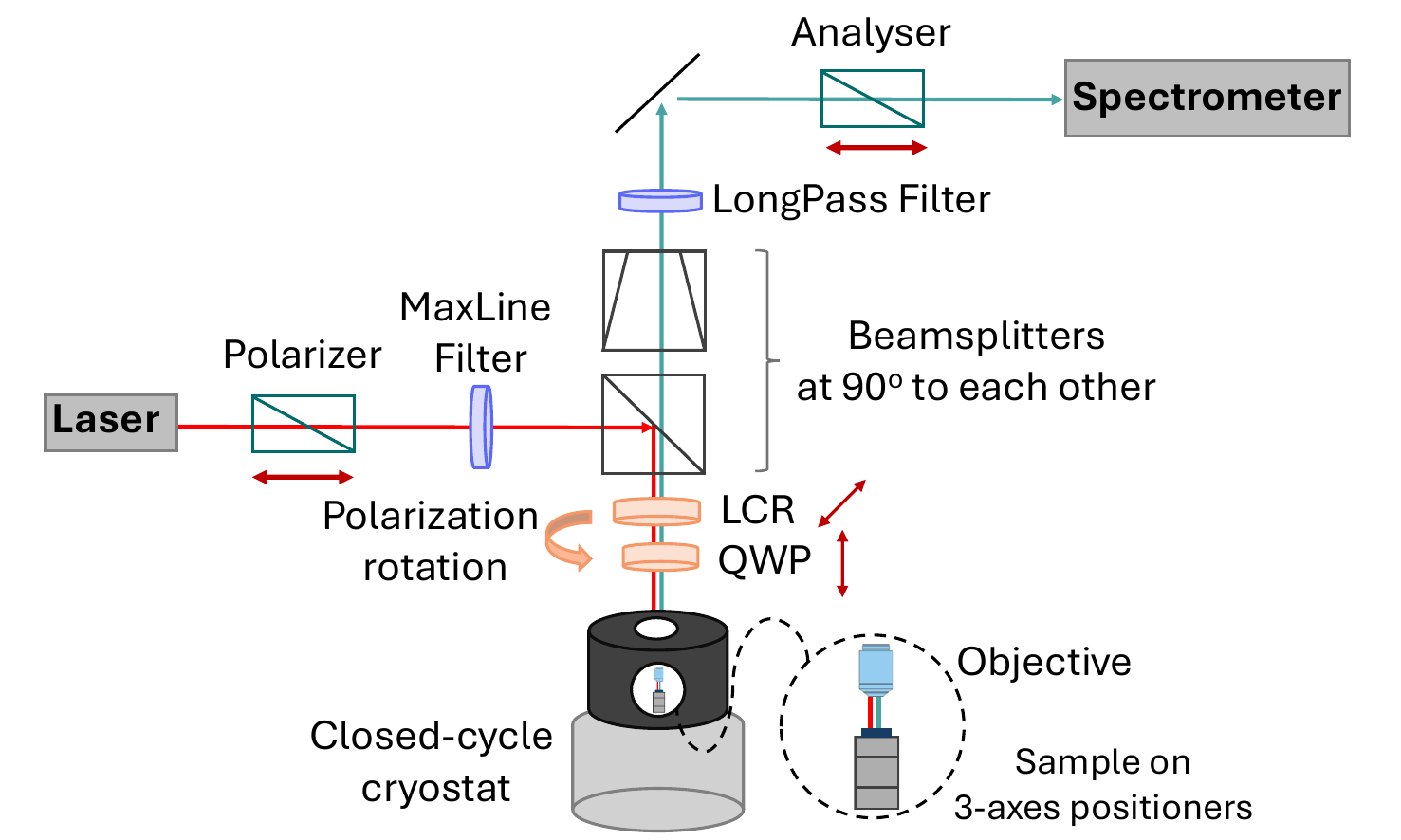}
\caption{\textbf{Experimental Raman scattering set-up.} LCR stands for liquid crystal retarder and QWP
for quarter-wave plate. Red arrows indicate the orientation of the optical axis of each polarization
element. For the polarizer and analyzer, the orientation is horizontal, providing a co-polarized measurement scheme. The LCR optical axis is at 45$^o$ to the incident polarization, and the QWP is perpendicular to the incidence.
}
\label{fig:fig1} 
\end{figure*}

\begin{figure*}
\includegraphics[width=1\linewidth]{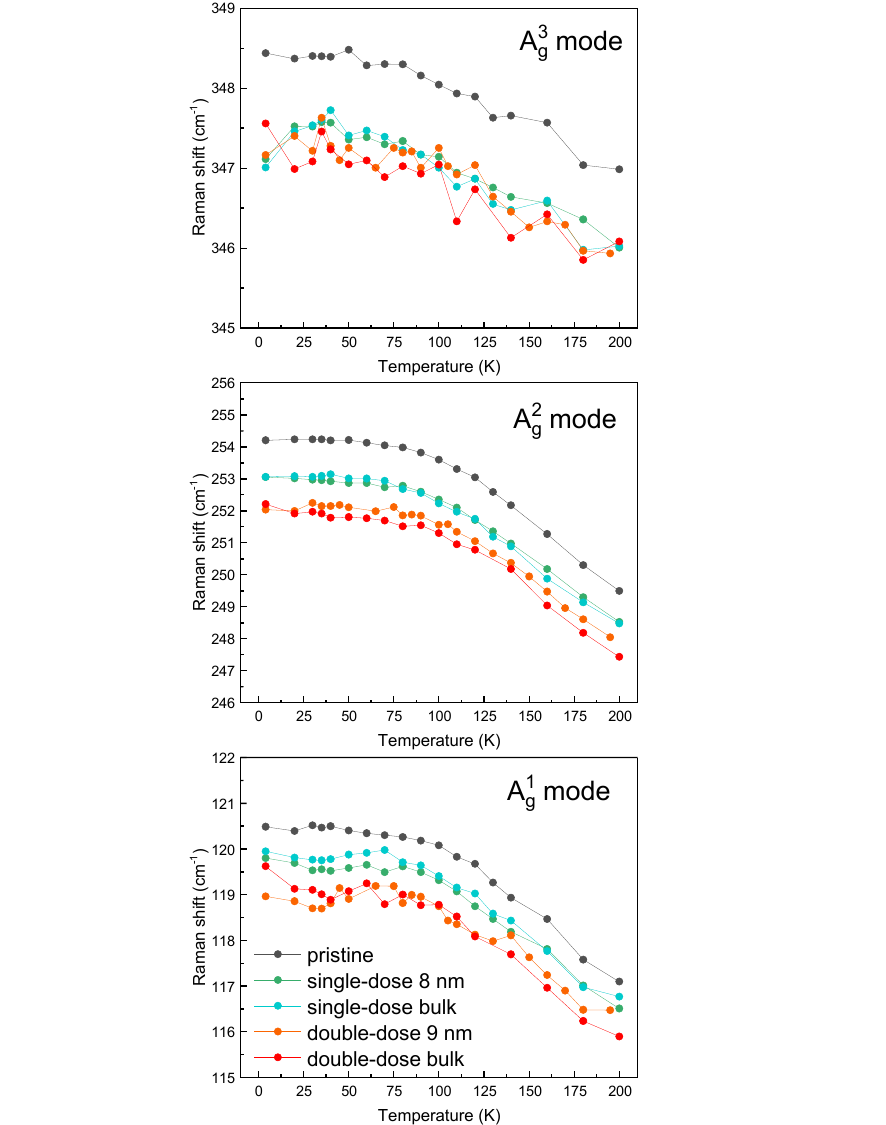}
\caption{\textbf{Temperature dependence of the Raman shift for pristine and irradiated samples.} The bottom, middle, and top panels show the temperature dependence of the Raman shift for the A$_g^1$, A$_g^2$, and A$_g^3$ phonon modes, respectively. Measurements were performed on pristine (black curves) and irradiated (colored curves) samples of different thicknesses under 1.96 eV excitation. The Raman peak positions were determined by Gaussian fits to the measured spectra.
}
\label{fig:fig1} 
\end{figure*}

\begin{figure*}
\includegraphics[width=1\linewidth]{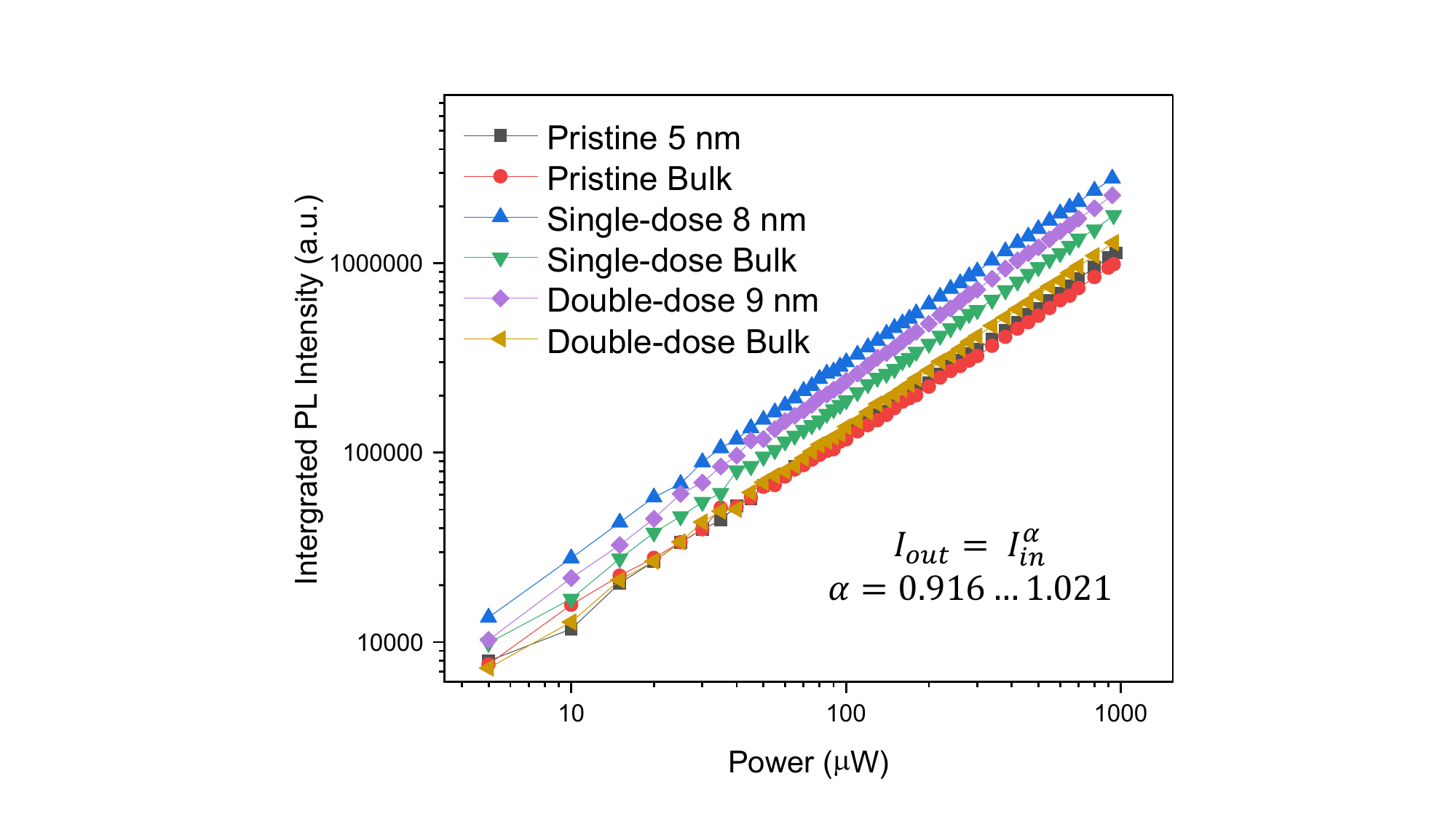}
\caption{\textbf{Excitation power dependence of the integrated PL intensity for pristine and irradiated samples.} Measurements were performed on pristine (black and red curves) and irradiated (colored curves) samples of different thicknesses under 1.96 eV excitation. The extracted power-law exponent, $\alpha$, ranges from 0.916 to 1.021 for all samples, confirming a linear dependence of the PL intensity on excitation power and indicating the absence of saturation up to an excitation power of 1000~$\mu$W.
}
\label{fig:fig1} 
\end{figure*}

\end{document}